\documentclass[prl,aps,twocolumn,showpacs]{revtex4}
\usepackage{epsfig}
\usepackage{bm}

\begin{document}

\title{Nonlocal-local multimode bifurcation in turbulence}
\author{\small  A. Bershadskii}
\affiliation{\small {ICAR, P.O. Box 31155, Jerusalem 91000, Israel }}

\begin{abstract}
It is shown that a mechanism of energy redistribution and dissipation by the inertial 
waves can be effectively utilized in isotropic turbulence at small Reynolds numbers. 
This mechanism totally suppresses the local interactions (cascades) in isotropic turbulence 
at the Taylor-scale based Reynolds number $R_{\lambda} < 10^2$. This value of $R_{\lambda}$ (that can be considered 
as a bifurcation value at which the 
local regime emerges from the nonlocal one in isotropic turbulence) is in agreement with recent 
direct numerical simulations data. Applicability of this approach to channel flows is also 
briefly discussed. A theory of multimode bifurcations has been developed in order to explain anomalous 
(in comparison with the Landau-Hopf bifurcations) properties of the nonlocal-local bifurcation 
in isotropic turbulence.
\end{abstract}


\maketitle

\section{Introduction}  

Thanks to increase of resolution of direct numerical simulations (DNS) 
the significant problem of emergence of the local interactions 
in isotropic turbulence became available for a quantitative 
investigation (see, for instance, Refs. \cite{gfn}-\cite{map} and references therein). 
Using the DNS data it is shown in a recent paper \cite{b1} that nonlocal 
interactions are completely dominating in {\it low} Reynolds number isotropic 
turbulence, as well as in a near dissipation range at moderate and large Reynolds 
numbers. The bottleneck phenomenon \cite{f},\cite{lm}, produced by these interactions, 
has an universal character. In corresponding near-dissipation 
range of scales \cite{b1}: $\eta < r < \eta_s$ 
the nonlocal interactions are completely dominating ones in isotropic turbulence 
(where $\eta_s \simeq 27 \eta$ and $\eta$ 
is the dissipation or Kolmogorov scale). Another scale: $r_c$, 
is a 'stability exchange"' scale \cite{b2}. Namely, for scales $r > r_c$, the
local (cascade) regime is stable and the nonlocal regime is unstable,
whereas for scales $r < r_c$ the local regime is unstable and the nonlocal
regime is stable. The normalized scale $\eta_s/\eta$ is constant \cite{b1} 
whereas $r_c/\eta$ increases with $R_{\lambda}$ \cite{b2}. Therefore, there is a value of 
$R_{\lambda}$ at which $r_c=\eta_s$ (see Figs. 1 and 2). This value of $R_{\lambda}$ can be 
considered as a bifurcation value of Reynolds number at which the local regime (still unstable) emerges from the 
nonlocal one. In present paper a phenomenological approach is suggested in order to compute 
bifurcation value of $R_{\lambda}$. 

There is an overlapping of the local and nonlocal regimes in the 
range of scales $r_c > r > \eta_s$ \cite{b1},\cite{b2}. Moreover, in the 
turbulent environment even the bifurcation scale $\eta_s$ might fluctuate 
around its mean-field value along with the dissipation scale $\eta$ 
(see \cite{ssy},\cite{b1},\cite{biferale}). Hence, at the nonlocal-local bifurcation 
the supercritical and subcritical secondary oscillations should co-exist. 
However, at the Landau-Hopf bifurcation this cannot be possible (see, 
for instance, Refs. \cite{ll},\cite{gh}). 
In order to explain this 'anomalous' behavior of the nonlocal-local bifurcation we will develop 
a theory of multimode bifurcations. In this theory several modes (which frequencies
are in certain specific relationships) are exited simultaneously. At a bifurcation in a turbulent environment 
at least second harmonic should be taken into account. From dynamical
point of view this situation is more complex than the one-mode (Landau-Hopf) bifurcation
and corresponding dynamical equations for the exited modes are multidimensional (in the
one-mode Landau-Hopf bifurcation the dynamical equation for the exited mode is two-dimensional 
in real variables). It is well known that behavior of dynamical systems with
dimension more than two is significantly different from the two-dimensional dynamical
systems. In particular, the former equations can generate stochastic attractors. This gives to the 
multimodal bifurcations considerable capacity in the phase space, which (unlike to the one-mode bifurcations) 
provide the multimode bifurcations a possibility to survive in turbulent environment.

\section{Phenomenology} 

A large vortex produces a Coriolis force field in the
domain of space that it occupies, which is due to the
rotational motion of the vortex. The comparatively slow
motion of a large vortex affords time for the process of
radiation of inertial waves to be realized by small-scale
fluctuations in the Coriolis force field of this vortex
\cite{it},\cite{hgm}. Such inertial waves transfer
the energy being emitted by the fluctuations to the viscous
layers on solid walls or to the viscous layers generated by the inertial 
waves on the effective boundaries of the large scale vortices \cite{phi}. 
It is known that the energy brought by the inertial waves to the viscous layers
dissipates effectively therein \cite{it},\cite{hgm},\cite{phi}. 
The dissipation layer of inertial waves has a thickness of order 
$$
l \propto (\nu / \Omega )^{1/2},    \eqno{(1)}
$$
where $~\nu~$ is the molecular viscosity, and $~\Omega~$ is a
characteristic angular velocity of the large-scale vortex. To
estimate the value of $~\Omega,~$ one can use 
$
\Omega \propto 2 \pi u_0/L. 
$
where $u_0$ is a typical large-scale velocity, and $L$ is a typical scale of 
the large-scale vortices.
Then 
$$
l \propto (\nu~L/2 \pi u_0)^{1/2}.  \eqno{(2)}
$$
If we define large-scale Reynolds number $R_L=Lu_0/\nu$, then we obtain
$$
l \propto \frac{L}{\sqrt{2 \pi} R_L^{1/2}}.  \eqno{(3)}
$$
Or, using the Taylor-scale based Reynolds number $R_{\lambda}\simeq 4(R_L)^{1/2}$ \cite{hin},\cite{l}, 
$$
l \propto \frac{4L}{\sqrt{2 \pi} R_{\lambda}} \eqno{(4)}
$$ 
The dissipative layer of the inertial waves is the 'friction' 
layer of the corresponding large-scale vortices. 

Part of the energy entrained by the inertial waves
 dissipates in the viscous layers while another part, being 
reflected, returns to the volume occupied by the bulk of the vortex.
 Decay of turbulent fluctuation kinetic
energy a result of its removal by inertial waves 
to the 'friction' layers can be described by equation \cite{it}:
$$
\frac{d\langle u^2 \rangle}{dt} = - \alpha 
\frac{(\nu \Omega)^{1/2}}{L}~\langle u^2 
\rangle,    \eqno{(5)}
$$
where ~$\alpha~$ is a dimensionless constant.
 If $~\Omega~$ and
$~L~$ are considered constants (or slowly changing), then it follows from (5):
$$
\langle u^2 \rangle \propto exp-\lambda t,    \eqno{(6)}
$$
where $~\lambda = \alpha (\nu \Omega)^{1/2} / L~$. 
This decay can be taken into account in the equation for the
spectral tensor in the "external friction" approximation 
\cite{it}:
$$
\frac{\partial E_{ij}}{\partial t} =  
T_{ij} (k,t) - \lambda E_{ij},   \eqno{(7)}
$$
where $~T_{ij}~$ characterizes the spectral  energy transfer
due to the nonlinear effects. The "external friction" reflects the
interaction between inertial effects and the viscous ('friction') layers, and it 
is a phenomenological substitute to the viscous term.
Since the inertial term in the Navier-Stokes equation is
quadratic in the velocity, then for small $~E_{ij}~$ the
functional $~T_{ij}~$ is a homogeneous functional 
of $~E_{ij}~$ of order 3/2. Let us make the substitution
$$
E_{ij}^* (k,t) = E_{ij} (k,t) e^{-\lambda t}.  \eqno{(8)}  
$$
Substituting (8) into (7) we obtain
$$
\frac{\partial E_{ij}^{*}}{\partial t} = T_{ij} (k,t)
 ~ e^{- \lambda t/2 }.    \eqno{(9)}
$$
Now, we make a replacement of the time
$$
t^{'} = \frac{2}{\lambda} (1 - e^{-\lambda t/2}).  \eqno{(10)}
$$
Then it follows from (9) that
$$
\frac{\partial E_{ij}^{*}}{\partial t^{'}} =
T_{ij}(k,t^{'}).  \eqno{(11)}
$$
Replacements (8) and (10) reduce the problem
with "external friction" to a problem without
friction (the initial conditions are evidently identical).
As $t \rightarrow \infty~$, it follows from (10) that $~t^{'}
 \rightarrow 2/\lambda,~$ so that the whole evolution with
"external friction", Eq. (7), is stacked in the interval $~(0,~
2/\lambda)~$ of the evolution described by ideal Eq. (11).

\begin{figure} \vspace{-0.5cm}\centering
\epsfig{width=.45\textwidth,file=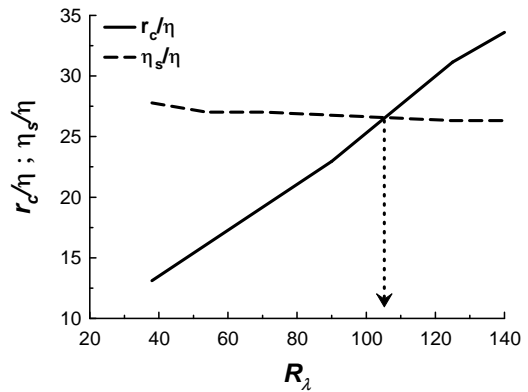} \vspace{-5cm}
\caption{Scales $r_c/\eta$ (solid line) \cite{b2} and $\eta_s/\eta$ \cite{b1} (dashed line) 
vs. $R_{\lambda}$. The arrow indicates the transitional value of the $R_{\lambda} \simeq 10^2$.}
\end{figure}

On the other hand, it is known (see, for instance, Refs. \cite{es},\cite{pg},\cite{pelz}) 
that there is a certain characteristic time of an inviscid cascade process development
$$
t_c \simeq 2 ~ \frac{L}{u_0}    \eqno{(12)}.
$$
Taking the preceding into account, an estimate can be made
of the condition under which the cascade process cannot be
developed successfully (because of the action of the "external
friction"). This condition has the form
$$
2/\lambda ~< t_c \simeq 2~ \frac{L}{u_0},  \eqno{(13)}
$$
or
$$ 
2L/\alpha (\nu \Omega )^{1/2} ~<~ 2~ \frac{L}{u_0}.  \eqno{(14)}
$$
Using estimate $~\alpha \simeq
10~$ \cite{it} we can rewrite condition (14) in the form
$$
R_{\lambda} < 10^2.  \eqno{(15)}
$$  

Thus, in isotropic turbulence with $R_{\lambda} < 10^2$ the cascade process (local interactions) 
cannot be developed successfully (cf. Ref. \cite{demo}). 
Figure 1 shows dependence of the scales $r_c/\eta$ \cite{b2} and $\eta_s/\eta$ \cite{b1} on 
$R_{\lambda}$. 
Intersection of the lines: $r_c/\eta$ vs. $R_{\lambda}$ and $\eta_s/\eta$ vs. $R_{\lambda}$, indicates the 
transitional value of $R_{\lambda} \simeq 10^2$ (see Eq. (15) and Introduction). Figure 2 illustrates 
the transition as it appears in the energy spectra of isotropic turbulence.
\begin{figure} \vspace{-0.3cm}\centering
\epsfig{width=.45\textwidth,file=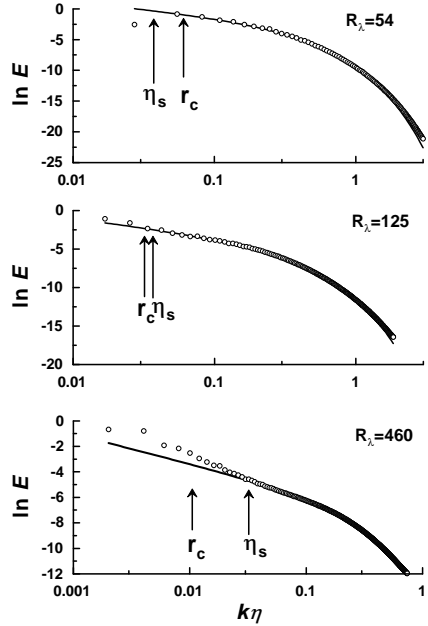} \vspace{-2.5cm}
\caption{Three-dimensional spectra from a DNS performed in \cite{gfn} for different Reynolds numbers. 
The solid curves in this figure correspond to the best fit by nonlocal interactions \cite{b1}.}
\end{figure}

\section{Separation}

One can also try to apply the above consideration to channel flows. 
The large-scale vortices acquire a metastable state in
channel turbulence because of the action of inertial forces. 
These forces are equilibrated by viscous friction on the solid
walls of the motion domain. The dissipative layer of the inertial waves 
on the solid walls is the 'friction' layer of the corresponding large-scale vortices. 
Therefore it is plausible to assume that $~l \simeq y_p$, where
$~y_p~$ is the characteristic size of the 'critical' layer \cite{sreeni},\cite{co}. 
With this assumption  we obtain from Eq. (3)
$$
y_{p}^{+} \propto R_{*}^{1/2},  \eqno{(16)}
$$
where $~y_{p}^{+} = y_p~u_* /\nu~$,  $~u_*~$ is so-called friction 
velocity, and $~R_* = L~u_*/ \nu $ \cite{sreeni} (here $L$ is a characteristic size of the channel, and 
$u_*\simeq u_0$). 
Eq. (16) is in agreement with the data presented in Ref. \cite{sreeni}. This can be 
considered as a positive indication of applicability of the above developed 
approach to the channel flows as well. However, in order to 
estimate transitional (from nonlocal to local turbulence) value of Reynolds number 
for the channel flows one needs in an analogy of the estimate (12) for such flows \cite{co}. 

\section{Multimode bifurcations}

Existence of the overlapping range of scales: $r_c > r > \eta_s$, implies co-existence 
of the supercritical and subcritical secondary oscillations at the nonlocal-local bifurcation 
in turbulent media. At the standard Landau-Hopf bifurcation \cite{ll},\cite{gh} such co-existence is impossible. 
Therefore, one needs in 
a generalization of the bifurcation theory (see Ref. \cite{ponty},\cite{lep} for other examples of 
bifurcations in turbulent and noisy media). The bifurcation problem for stationary solution 
of an evolutionary nonlinear equation in a Hilbert space generally can be reduced to investigation of
instability of trivial solution ($u=0$) of an equation 
$$
\frac{du}{dt} =\hat{L}(u) + \sum_{n=2}^{N} \hat{F}_n (u) \eqno{(17)} 
$$
where $u$ is an element of some Hilbert space, 
$\hat{L}$ is a linear operator in this space, $\hat{F}_n (u)$ are nonlinear operators 
in this space such that $\hat{F}_n(\lambda u) = \lambda^n \hat{F}_n (u)$, and $N$ is 
called order of nonlinearity. Generally the linear operator $\hat{L}$ is a 
nonselfadjoint operator. Complete system of
vectors in the Hilbert space contains eigenstates and associated eigenstates 
of this operator \cite{gk}. Let us expand a solution of the equation (17) in
series of this full system of the vectors 
$$
u(t) = \sum_{k=1}^{\infty} a_k (t) \psi_k \eqno{(18)} 
$$
Since physical fields can take only real values the spectrum of operator $\hat{L}$
always contains pairs of complex-conjugate eigenvalues \cite{gk}: $\hat{L} \psi =
\gamma \psi$ and $\hat{L}\overline{\psi} = \overline{\gamma} \overline{\psi}$, where 
$\overline{\gamma}$ is complex conjugate to $\gamma$. 

Let us start from the one-mode case. Let to a pair of such eigenvalues: 
$\gamma_1 = \delta + i\omega_0$, and $\overline{\gamma_1}$ (corresponding to the 
pair of eigenstates $a_1$ and $a_2$) be located in a
straight line which placed in a small vicinity of the imaginary axis
(i.e. $1 \gg |\delta|$). Rest of the eigenvalues of the operator $\hat{L}$ are
located on the left side of this straight line and of the imaginary 
axis. The parameter $\delta$ is control parameter of the system 
($\delta \sim (R_{L}-R_{L,c})$, and $R_{L,c}$ is a bifurcation value of the large-scale 
Reynolds number).  For small enough amplitude of the oscillations 
we can restrict ourselves by consideration of $N=2$. Let $\gamma_1$ be 
simple eigenvalues \cite{gk}. Then, substituting the series (18) into (17) 
and multiplying on the eigenvectors of the operator $\hat{L}^{+}$ (adjoint to $\hat{L}$) 
in the Hilbert space we obtain for $a_1$ equation: 
$$
\frac{da_1}{dt} = \gamma_1 a_j + \sum_{k,m=1}^{\infty} b_{km} a_ka_m,~~~  
  \eqno{(19)}
$$
$$
b_{km} = (\phi_1,F_2 (\psi_k,\psi_m))/(\phi_1,\psi_1);
$$
where $\phi_j$ are eigenvectors of the operator $L^{+}$. We have taken 
into account that the vector $\phi_1$ is orthogonal to all 
$\psi_k$ excluding $\psi_1$. Equation for $a_2 = \overline{a_1}$ can 
be obtained using the complex conjugate operation.

Let us introduce the variable $\Theta = \omega t$, where $\omega$ is
frequency of the auto-oscillations. Let us expend $a_n (\Theta)$ and $\omega$
in analytic series on $\delta$ 
$$
a_n = \delta a_n^{(1)} + \delta^2 a_n^{(2)} + ...  \eqno{(20)}
$$
$$
\omega = \omega_0 + \delta \omega_1 + \delta^2 \omega_2 +... \eqno{(21)}
$$
Substituting these expansions into equations for $a_n$ we obtain in the first
order 
$$
a_1^{(1)} (\Theta) = A_1 \exp (i\Theta) \eqno{(22)} 
$$
Amplitude $A_1$ can be obtained from next orders of the expansion. It can be
shown that for $n > 2$ 
$$
a_n^{(1)}(\Theta) = (c_0 +c_1\Theta + ... + c_k \Theta^k) \exp (\gamma_n 
\frac{\Theta}{\omega_0})   \eqno{(23)} 
$$
(the polynomial multiplier is related to the associated eigenvectors 
\cite{gk}).
Because $\Re\{\gamma_n \} < 0$ for $n > 2$, the $a_n^{(1)} (\Theta)$ decay
with time for $n > 2$. Further the terms decaying with time will not be
taken into account.

In the second order we obtain 
$$
\frac{da_1^{(2)}}{d\Theta} - ia_1^{(2)} 
= -\frac{\omega_1}{\omega_0} 
\frac{da_1^{(1)}}{d\Theta} +\frac{\beta}{\omega_0} a_1^{(1)} +
\frac{1}{\omega_0}\sum_{n,m=1}^{2} b_{nm}^{(0)} a_n^{(1)}a_m^{(1)} 
   \eqno{(24)}
$$
where $b_{km}^{(0)} = b_{km}(\delta = 0)$, $\beta = 1$ for supercritical
and $\beta =-1$ for subcritical cases. Analogous equation can be obtained 
for the complex conjugate function $a_2^{(2)}$ by complex conjugation operation.

If one substitutes $a_1^{(1)}$ from (22) into right-hand side of 
Eq. (24), then terms $\sim \exp (i\Theta)$ and $\sim \exp (-i\Theta )$ will
lead to resonance, which cannot be eliminated due to the necessary 
for this elimination condition 
$$
\beta - i\omega_1 =0     \eqno{(25)}
$$
cannot be satisfied. Therefore the analytical series expansions (20),(21) 
are not applicable in the one-mode case. 

  Instead of the analytic expansions (20),(21) one can use non-analytical (Hopf) 
series on $\delta^{1/2}$
$$
a_n = \delta^{1/2} a_n^{(1)} + \delta a_n^{(2)} + \delta^{(3/2)} a_n^{(3)} 
+...    \eqno{(26)}
$$
$$
\omega = \omega_0 + \delta^{1/2} \omega_1 + \delta \omega_2 + ... 
  \eqno{(27)} 
$$
In the first order we again obtain for $a_1^{(1)}$ representation (22). But 
in the next order we now obtain equation  
$$
\frac{da_1^{(2)}}{d\Theta} - ia_1^{(2)} 
= -\frac{\omega_1}{\omega_0} 
\frac{da_1^{(1)}}{d\Theta}  + \frac{1}{\omega_0} 
\sum_{n,m=1}^2 b_{nm}^{(0)} a_n^{(1)}a_m^{(1)}            \eqno{(28)} 
$$  
The resonance term can be now eliminated by choosing $\omega_1=0$. 
Analogous situation takes place for complex conjugated equation 
for $a_2^{(2)}$. Then in the next order 
$$
\frac{da_1^{(3)}}{d\Theta} -i a_1^{(3)} = -\frac{\omega_2}{\omega_0} 
\frac{da_1^{(1)}}{d\Theta} + \frac{1}{\omega_0} a_1^{(1)} + 
$$
$$
\frac{1}{\omega_0} \sum_{n,m=1}^2 b_{nm}^{(0)} [a_n^{(1)}a_m^{(2)} + 
a_n^{(2)}a_m^{(1)}]  \eqno{(29)}
$$

Since the last term in equation (28) gives terms 
$\sim \exp(i2\Theta)$ and $\sim \exp-(i2\Theta)$ into its solution 
the resonance terms can now be eliminated from the 
equation (29) and conditions of this elimination provide us 
$A_1$ and $\omega_2$. This is {\it Hopf} bifurcation which gives 
dependence $a_1 \sim \delta^{1/2}$.

   If the original stationary
solution has groups of symmetry, then more than one pair of the eigenvalues
can be located in a straight line parallel to the imaginary axis. Let to two
pairs of such eigenvalues: $\gamma_1 = \delta + i\omega_0$, $\overline{\gamma_1}$
and $\gamma_2 = \delta + 2i\omega_0$, $\overline{\gamma_2}$, be located in a
straight line which placed in a small vicinity of of the imaginary axis. 
Rest of the eigenvalues of the operator $\hat{L}$ are
located on the left side of this straight line and of the imaginary 
axis. It
will be clear from further considerations that fulfillment of these conditions
with accuracy $O(\delta^3)$ is enough for existence of the auto-oscillations
in the system.
 Let $\gamma_1$ and $\gamma_2$ be simple eigenvalues \cite
{gk}. Then, substituting the series (18) into (17) and multiplying on the
eigenvectors of the operator $\hat{L}^{+}$ (adjoint to $\hat{L}$) in the Hilbert space
we obtain for $a_1$ and $a_2$ following equations: 
$$
\frac{da_j}{dt} = \gamma_j a_j + \sum_{k,m=1}^{\infty} b_{jkm} a_ka_m,~~~ j=1,2 
   \eqno{(30)}
$$
$$
b_{jkm} = (\phi_j,F_2 (\psi_k,\psi_m))/(\phi_j,\psi_j);
$$
where $\phi_j$ are eigenvectors of the operator $\hat{L}^{+}$. We have taken 
into account that the vector $\phi_j$ is orthogonal to all $%
\psi_k$ excluding $\psi_1$, and $\phi_2$ is orthogonal to all $\psi_k$
excluding $\psi_2$. Equations for $a_3 = \overline{a_1}$ and for $a_4= 
\overline{a_2}$ can be obtained using the complex conjugate operation.

Let us now expend $a_n (\Theta)$ and $\omega$ in the analytic series (20),(21). 
Substituting these expansions into equations for $a_n$ we obtain in the first
order 
$$
a_j^{(1)} (\Theta) = A_j \exp (i\frac{\Im \{ \gamma_j \}}{\omega_0} \Theta) 
   \eqno{(31)} 
$$
Amplitude $A_j$ can be obtained from next orders of the expansion. 
In the second order we obtain 
$$
\frac{da_j^{(2)}}{d\Theta} - i\frac{\Im \{\gamma_j \}}{\omega_0} a_j^{(2)} 
= -\frac{\omega_1}{\omega_0} 
\frac{da_j^{(1)}}{d\Theta} +\frac{\beta}{\omega_0} a_j^{(1)} +
$$
$$
\frac{1}{\omega_0} \sum_{n,m=1}^{4} b_{jnm}^{(0)} a_n^{(1)}a_m^{(1)}   
   \eqno{(32)} 
$$
If one substitutes $a_j^{(1)}$ taken from (31) into right hand side of 
equations (32), then terms $\sim \exp (i\Theta)$ and 
$\sim \exp (i2\Theta )$ will
lead to resonances for $j =1$ and for $j=2$ respectively. If we write
conditions of suppression of these resonances 
$$
A_2 = \frac{b_{211}^{(0)} A_1^2}{(i2\omega_1 - \beta)\omega_0} 
$$
$$
(i2\omega_1 -\beta) (\beta -i\omega_1) + |A_1|^2k =0 
$$
$$
k=k_1+ik_2= \frac{1}{\omega_0}\Re\{(b_{123}^{(0)} + b_{132}^{(0)})
b_{211}^{(0)}\} + i \Im\{b_{123}^{(0)} + b_{132}^{(0)}\} 
$$
we obtain 
$$
|A_1|^2 = -\frac{3\beta}{k_2} \omega_1, \eqno{(33)} 
$$
$$
\omega_1 = \beta \left[ \frac{3k_1}{4k_2} \pm 
\frac{|\beta |}{\beta}\sqrt{\frac{9k_1^2}{16k_2^2} + \frac{1}{2}} \right]. 
  \eqno{(34)} 
$$
We have found only modulus of $A_1$, but we can let $A_1$ be real due to 
initial phase of the two-mode auto-oscillations can be arbitrary chosen.

    Thus one can see that for the multimode bifurcation the amplitude of the
secondary oscillations is proportional to the control parameter, $\delta$, 
while for one-mode (Landau-Hopf) bifurcation this amplitude is 
proportional to $\sqrt{\delta}$. It is clear that 
the anomalous dependence will also take place in, for example, more complex 
situation of three complex conjugated pairs with frequencies $\omega_0$, 
$(n-1)\omega_0$, and $n \omega_0$ (where $n$ is an arbitrary integer). 
One can construct other combinations of the frequencies 
leading to the anomalous dependence. 

In our case $\delta \sim (R_L-R_{L,c})$ (where $R_{L,c}$ is a bifurcation value 
of the large-scale Reynolds number). Let us denote for longitudinal fluctuations of the 
velocity field $v$ a rms-value: $u'=\langle v^2 \rangle^{1/2}$. Then for the multimode 
bifurcation 
$$
u'= u_c^{'}+ C (R_L-R_{L,c})+... , \eqno{(35)}
$$
i.e
$$
u'\simeq u_c^{'}+\frac{1}{16}C (R_{\lambda}^2- R_{\lambda , c}^2) + ... \eqno{(36)}
$$
(where $C$ is certain constant). Hence, for the multimode bifurcation $u'$ is 
a linear function of $R_{\lambda}^2$ in a vicinity of the bifurcation value $R_{\lambda ,c}$ 
(for the standard one-mode Landau-Hopf bifurcation \cite{ll},\cite{gh} we have $u'=u_c^{'}+ 
C\sqrt{R_L-R_{L,c}}$, instead of the Eq. (35)). 
Figure 3 shows dependence $u'$ on $R_{\lambda}^2$ for the data obtained in a grid-turbulence 
laboratory experiment \cite{gm}. The straight line is drawn in this figure in order 
to indicate agreement with the Eq. (36) (i.e. with the multimode bifurcation prediction).
\begin{figure} \vspace{-0.5cm}\centering
\epsfig{width=.45\textwidth,file=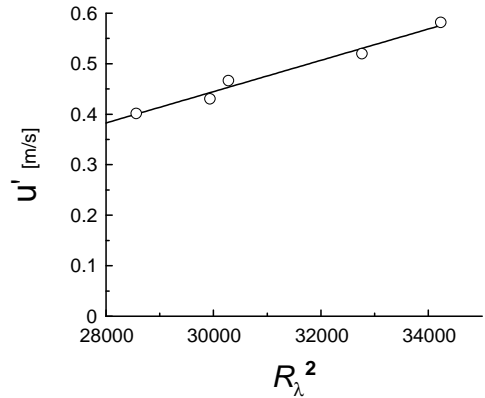} \vspace{-4.5cm}
\caption{Dependence of $u'$ on $R_{\lambda}^2$ for a grid-turbulence 
experiment \cite{gm}. The straight line indicates Eq. (36).}
\end{figure}

\section{Stability } 

It follows from Eq. (33) that its right hand side should be positive. It can be
obtained for any sign of $\beta$ if one chooses an appropriate sign in (34).
Therefore, the supercritical and subcritical secondary oscillations co-exist
at the multimode bifurcation (while at the standard one-mode Landau-Hopf 
bifurcations this cannot be possible). This co-existence makes the question 
about stability of the secondary regimes rather important.

Let us consider the evolutionary equation linearized on the secondary regime
$$
(\omega _{0}+\delta \omega )\frac{da_{j}}{d\Theta }=(\beta \delta +\gamma
_{j}\omega _{0})a_{j}+
$$
$$
+ \delta \sum_{k,m=1}^{4}b_{jkm}(a_{k}^{(1)}a_{m}+a_{m}^{(1)}a_{k}) \eqno{(37)}
$$
where $a_{k}^{(1)}(\Theta )$ are the first order (on the parameter $\delta$) 
modes of the considered secondary oscillations. The Floquet representation
for evolutionary operator of these equations is \cite{dk}
$$
\hat{V}(\Theta )=\hat{Q}(\Theta )\exp (\hat{G}\Theta ).         \eqno{(38)}
$$
Stability or instability of the secondary oscillation is determined by
spectrum of the operator $\hat{G}$. We use small parameter $\delta $ to find when
the spectrum of the operator $\hat{G}$ has its points located in the right
semi-plane. Let us expend operator $\hat{G}$ 
$$
\hat{G}=\hat{G}_{0}+\delta \hat{G}_{1}+...        \eqno{(39)}
$$
where 
$$
\hat{G}_{0}=i~diag\{1,2\}    \eqno{(40)}
$$
Eigenvalues of the operator $\hat{G}_{0}$ are: $\sigma _{1}^{(0)}=i$ and $\sigma
_{2}^{(0)}=2i$. Corresponding eigenvectors are: $x_{1}=(1,0)$, $x_{2}=(0,1)$%
. To find operator $\hat{G}_{1}$ we use the Krein representation \cite{dk} 
$$
\hat{G}_{1}=\frac{1}{4\pi ^{2}}\int_0^{T_0} \oint_{\Gamma_0} \oint_{\Gamma_0} 
\frac{(\lambda -\mu )e^{(\mu
-\lambda )\tau }}{1-e^{(\mu -\lambda )\tau }} \times
$$
$$
\times (\hat{G}_{0}-\lambda
\hat{I})^{-1}\hat{R}_{1}(\hat{G}_{0}-\mu \hat{I})^{-1}d\tau d\lambda d\mu \eqno{(41)}
$$
where $T_{0}=2\pi \omega _{0}^{-1}$, 
$$
\hat{R}_{1}=\omega _{0}^{-1}diag\{\beta -i\omega _{1},~~~\beta -2i\omega _{1}\}
$$
and $\Gamma _{0}$ is a contour enveloping the spectrum of the operator $\hat{G}_{0}
$. Harmonic perturbations of the right hand side of the evolutionary equation
do not give a contribution to the $\Re \{\sigma \} \sim \delta$. If we now 
expand spectrum of the operator $\hat{G}$: $\sigma = \sigma^{(0)} + \delta
\sigma^{(1)} + ...$, then we obtain 
$$
\sigma^{(1)} = (y, \hat{G}_1x) \eqno{(42)} 
$$
where $x$ is eigenvalue of the operator $\hat{G}_0$, $y$ is eigenvalue of the
operator conjugate to $\hat{G}_0$ and $(,)$ means scalar multiplication. Then,
substituting $\hat{G}_1$ from (41) into (42) and calculating the integrals we
obtain 
$$
\Re \{ \sigma \} = - \beta \omega_0^{-1} \delta + ...  \eqno{(43)} 
$$
Therefore, for small $\delta$ the subcritical secondary oscillations $(\beta
= -1)$ are unstable ($\Re \{\sigma \} > 0)$, whereas supercritical ($\beta =
1$) secondary oscillations are stable ($\Re \{ \sigma \} < 0$). 

While one-mode equation (19) is two-dimensional (for real variables) 
the multimode equations are multidimensional: two-mode equation 
has dimensionality  4, and the three-mode equation for the modes with 
$\omega_0$, $(n-1)\omega_0$ and $n\omega_0$ has dimensionality 6. It is 
well known that nonlinear equations with dimensionality larger than 2 and 
equations with dimensionality smaller (or equal) than 2 have very different 
properties. In particular, the former equations can generate stochastic 
attractors. These attractors have considerable capacity in the phase space 
that allows them to survive in the turbulent environment.

\acknowledgments

I thank T. Nakano, D. Fukayama, T. Gotoh,
and K. R. Sreenivasan for sharing their data and discussions.


\begin{thebibliography}{99}
\bibitem{gfn} T. Gotoh, D. Fukayama, and T. Nakano, Phys. Fluids {\bf 14} (2002) 1065.
\bibitem{ishi} T. Ishihara, Y. Kaneda, M. Yokokawa, K. Itakura, and A. Uno, 
J. Phys. Soc. Jpn. {\bf 74} (2005) 1464 .
\bibitem{ssy} J. Schumacher, K. R. Sreenivasan, and V. Yakhot, New Journal of Physics {\bf 9} (2007) 89.
\bibitem{map}  P. Mininni, A. Alexakis, A. Pouquet, Phys. Rev. E, 
{\bf 77} (2008) 036306. 
\bibitem{b1} A. Bershadskii, Phys. Fluids, {\bf 20} (2008) 085103.
\bibitem{f} G. Falkovich, Phys. Fluids, {\bf 6} (1994) 1411. 
\bibitem{lm} D. Lohse and A. Muller-Groeling, Phys. Rev. Lett. {\bf 74} (1995) 1747.
\bibitem{b2} A. Bershadskii, J. Stat. Phys.,  {\bf 128} (2007) 721.
\bibitem{biferale} L. Biferale, Phys. Fluids,   {\bf 20} (2008) 031703.
\bibitem{ll} L.D. Landau and E.M. Lifshitz, Fluid mechanics (Pergamon, Oxford, 1987). 
\bibitem{gh} J. Guckenheimer and P. Holmes, Nonlinear Oscillations, Dynamical Systems and
Bifurcations of Vector Fields (Springer-Verlag, NY, 1983).
\bibitem{it} A. Ibetson and D.J. Tritton, J.Fluid Mech., {\bf 68} (1975)
639.
\bibitem{hgm} E.J. Hopfinger, R.W. Griffits and M. Mory, J. de Mec., {\bf 2} (1983) 21.
\bibitem{phi} O.M. Phillips, Phys. Fluids, {\bf 6} (1963) 513. 
\bibitem{hin} J.O. Hinze, Turbulence, (McGraw-Hill, 1959).
\bibitem{l} D. Lohse, Phys. Rev. Lett., {\bf 73} (1994) 3223.
\bibitem{es} G.L. Eyink, and K.R. Sreenivasan, Rev. Mod. Phys., {\bf 78} (2006) 87.
\bibitem{pg} R.B. Pelz, and Y. Gulak, Phys. Rev. Lett. {\bf 25} (1997) 4998.
\bibitem{pelz} R.B. Pelz, Fluid Dynam. Res., {\bf 33} (2003) 207. 
\bibitem{demo} P.E. Dimotakis, J. Fluid Mech. {\bf 409} (2000) 69.
\bibitem{sreeni} K.R. Sreenivasan, A unified view of the origin and morphology of the turbulent
boundary layer structure, In "Turbulence Management and Relaminarization", 37-61, 
(ed. H.W. Leipman, R. Narasimha, Springer-Verlag Berlin, 1988).
\bibitem{co} S.J. Cowley, Laminar Boundary-Layer Theory:
A 20TH century paradox? in Proceedings of ICTAM 2000, eds. H. Aref and J.W. Phillips, 389-411, Kluwer (2001). 
\bibitem{ponty} Y. Ponty, J.-P. Laval, B. Dubrulle, F. Daviaud, and J.-F. Pinton, Phys. Rev. Lett., {\bf 99} 
224501 (2007).
\bibitem{lep} N. Leprovosta and B. Dubrulle, Eur. Phys. J. B, {\bf 44}, 395 (2005).
\bibitem{gk} I. Gochberg and M.G. Krein, Introduction to the theory of linear 
nonselfadjoint operators (Am. Math. Soc., NY, 1969).
\bibitem{gm} S. Cerutti and C. Meneveau, Phys. Fluids, {\bf 12} (2000) 1143.
\bibitem{dk} Yu.L. Daletskii and M.G. Krein, Stability of solutions of 
differential equations in Banach space (Am. Math. Soc., NY, 1974).
\end{thebibliography}
\end{document}